# Field-induced spin-density wave beyond hidden order in URu$_2$Si$_2$

W. Knafo[1], F. Duc[1], F. Bourdarot[2], K. Kuwahara[3], H. Nojiri[4], D. Aoki[5,6], J. Billette[1],

P. Frings[1], X. Tonon[7], E. Lelièvre-Berna[7], J. Flouquet[6], and L.-P. Regnault[7]

[1] *Laboratoire National des Champs Magnétiques Intenses, UPR 3228, CNRS-UPS-INSA-UGA, 143 Avenue de Rangueil, 31400 Toulouse, France.*

[2] *Service de Modélisation et d'Exploration des Matériaux, Université Grenoble Alpes et Commissariat à l'Energie Atomique, INAC, 17 rue des Martyrs, 38054 Grenoble, France.*

[3] *Institute of Quantum Beam Science, Ibaraki University, Mito 310-8512, Japan.*

[4] *Institute for Materials Research, Tohoku University, Sendai 980-8578, Japan.*

[5] *Institute for Materials Research, Tohoku University, Ibaraki 311-1313, Japan.*

[6] *Service Photonique, Electronique et Ingénierie Quantiques, Université Grenoble Alpes et Commissariat à l'Energie Atomique, INAC, 17 rue des Martyrs, 38054 Grenoble, France.*

[7] *Institut Laue-Langevin, 71 Avenue des Martyrs, CS 20156, 38042 Grenoble, France.*

*Correspondence and requests for materials should be addressed to W.K.*
*(email: william.knafo@lncmi.cnrs.fr)*

**$URu_2Si_2$ is one of the most enigmatic strongly-correlated-electron systems and offers a fertile testing ground for new concepts in condensed matter science. In spite of >30 years of intense research, no consensus on the order parameter of its low-temperature 'hidden-order' phase exists. A strong magnetic field transforms the hidden order into magnetically-ordered phases, whose order parameter has also been defying experimental observation. Here, thanks to an instrumentation breakthrough in high-field neutron scattering, we identify the field-induced phases of $URu_2Si_2$ as a spin-density-wave state with wavevector $k_1 = (0.6\ 0\ 0)$. The transition to the spin-density wave represents a unique touchstone for understanding the hidden-order phase. An intimate relationship between this magnetic structure, the magnetic fluctuations, and the Fermi surface is emphasized, calling for dedicated band structure calculations.**

Since three decades, a huge experimental and theoretical effort has been employed to try to determine the ground state of the heavy-fermion material $URu_2Si_2$, where a second-order phase transition occurs at the temperature $T_0 = 17.5$ K and for which the order parameter has resisted experimental identification. [1]. The case of $URu_2Si_2$ - which is also superconducting [2] - is not isolated, since mysterious electronic phases, identified as 'pseudo-gap' and 'nematic' phases, have also been reported in high-temperature cuprate and iron-based superconducting materials [3],[4]. Initially, a small magnetic moment of 0.02 $\mu_B$/U antiferromagnetically aligned along **c** and modulated with the wavevector $k_0 = (0\ 0\ 1)$ has been considered as the order parameter of $URu_2Si_2$ below $T_0$ [5]. However, this moment is too small to account for the large entropy associated with the transition and is now suspected to result from sample inhomogeneities [6]. In the hidden-order phase, the low-energy magnetic fluctuations are peaked at the hot wavevectors $k_0$ and $k_1 = (0.6\ 0\ 0)$ [7],[8], and the Fermi surface presents nesting with these two wavevectors [9],[10],[11],[12],[13],[14],[15] suggesting an itinerant origin of the magnetic properties. Recent experimental works have been interpreted as the signature of nematicity, that is, a correlated electronic state associated with a breaking of the four-fold symmetry, in the hidden-order phase [16],[17],[18].

The application of extreme conditions permits to modify the ground state of a system and get new insights about it. In URu$_2$Si$_2$ under pressure, the hidden-order state is replaced by an antiferromagnetic phase [19], whose order parameter is a moment of $\simeq$ 0.3-0.4 $\mu_B$/U aligned antiferromagnetically along **c** with the wavevector **k**$_0$ [20],[21]. From the similar Fermi surfaces observed in the hidden-order and pressure-driven antiferromagnetic phases, it has been proposed that the hidden-order, although of unknown nature, has a periodicity of wavevector **k**$_0$ [19]. Under a high magnetic field applied along the easy magnetic axis **c**, a cascade of first-order phase transitions at the fields $\mu_0H_1$ = 35 T, $\mu_0H_2$ = 36 / 37 T (rising / falling fields), and $\mu_0H_3$ = 39 T, leads to a polarized paramagnetic regime above $H_3$, where a magnetization of $\approx$ 1.4 $\mu_B$/U is reached (see Figure 1b) [22],[23]. Quantum oscillations probes have shown that a magnetic field along **c** induces Fermi surface instabilities deep inside the hidden-order phase, as well as at the magnetic transitions $H_1$, $H_2$, and $H_3$ [24],[25],[26],[27],[28]. By comparison with 3$d$ itinerant magnets, a novelty in heavy-fermion compounds is that a large magnetic polarization can be induced by a magnetic field, opening the path to new phases driven by combined changes of magnetic order and Fermi surface. Such interplay of metamagnetic and Lifshitz transitions has been evidenced in the heavy-fermion paramagnet CeRu$_2$Si$_2$ [29].

In Rh-doped U(Ru$_{0.96}$Rh$_{0.04}$)$_2$Si$_2$, no hidden-order phase is reported, but first-order transitions are also induced at magnetic fields of 26 and 37 T applied along **c** [30],[31],[32]. Between 26 and 37 T, a squared *'up-up-down'* ferrimagnetic phase was deduced from Bragg peaks (harmonics) at the wavevector **k**$_2$ = (2/3 0 0) and at the Brillouin zone center **k**$_{ZC}$ = (0 0 0) [33]. However, the extrapolation to the pure case of URu$_2$Si$_2$ is not straightforward, as doping may modify the low carrier content of the pure lattice, and, thus, the Fermi surface topology, leading to different nesting effects. Here, thanks to an experimental breakthrough allowing neutron diffraction up to 40 T, we have determined the magnetic structure of URu$_2$Si$_2$ in fields between 35 and 39 T. A tilt by 4.2 ° of the field direction from **c** in the (**c**,**a**) plane allowed accessing a large number of momentum transfers **Q,** with almost no modification of the magnetic phase diagram expected for **H** || **c** [26],[34]. The relationship between the high-field magnetic structure, the low-energy magnetic fluctuations, and the

Fermi surface is emphasized, appealing for new band structure calculations describing the electronic properties of URu$_2$Si$_2$.

## Results

**Spin-density wave with wavevector k$_1$.** Figure 1a shows the time profile of a magnetic field pulsed up to 38 T, as well as the time dependence of the diffracted neutron intensities recorded during 38-T pulses at the temperatures $T = 2$ and 18 K and at **Q** = (0.6 0 0) and (1.6 0 -1). These momentum transfers are satellites of wavevector **k**$_1$ = (0.6 0 0) = **Q** – **τ** around the structural Bragg positions **τ** = (0 0 0) and (1 0 -1), respectively. The high-field enhancement of the diffracted intensity at $T = 2$ K, absent at $T = 18$ K, shows that long-range ordering with wavevector **k**$_1$ = (0.6 0 0) is established at high field and low temperature. Figure 1c emphasizes in the window 32-41 T the field-dependence of the diffracted neutron intensities measured up to 40.5 T. At $T = 2$ K, the diffracted intensity with wavevector **k**$_1$ is enhanced in fields higher than $\mu_0 H_1 = 35$ T and drops down to the background level in fields higher than $\mu_0 H_3 = 39$ T. No hysteresis is observed at $H_1$ and $H_3$ within the neutron counting statistics, and a slight field-induced decrease of intensity, suggesting a modification of magnetic ordering, is visible at $\mu_0 H_2 \approx 36$ T for rising field and at $\mu_0 H_2 \approx 37$ T for falling field, in agreement with the hysteresis observed by magnetization [33] (Figure 1b). Figure 2 presents the high-field neutron diffracted intensity at momentum transfers **Q** in the vicinity of (0.6 0 0) at $T = 2$ K. The field-dependence of the intensity at momentum transfers ($Q_h$ 0 0) and (0.6 0 $Q_l$) is shown in fields up to 36.2 T in Figures 2a and 2b, respectively. Figures 2c and 2d present $Q_h$ - and $Q_l$ - scans at $\mu_0 H = 36$ T, confirming that the field-induced magnetic intensity is peaked at **Q** = (0.6 0 0). From the intensity of the magnetic Bragg peaks induced at **Q** = (0.6 0 0) and **Q** = (1.6 0 -1) at $\mu_0 H = 36$ T $> \mu_0 H_1$, we extract the amplitude $2M(\mathbf{k}_1) \simeq 0.5 \pm 0.05$ $\mu_B$/U of the sine-modulation with wavevector **k**$_1$ of the magnetic moment $M(\mathbf{k}_1,\mathbf{r})$ = $2M(\mathbf{k}_1)\cos(2\pi \mathbf{k}_1 \cdot \mathbf{r} + \varphi_1)$, where **r** is an U-ion position and $\varphi_1$ the associated phase (cf. Supplementary Information). This result is in agreement with the high-field magnetization [22], since the amplitude $2M(\mathbf{k}_1)$ extracted here at 36 T is of same order than the variation of magnetization $\Delta M \simeq 0.4$ $\mu_B$/U (estimated from 36 T to 39 T) induced by the quench of the field-induced ordered moments into a polarized regime above $\mu_0 H_3$ = 39 T.

The field-dependence of the diffracted intensity is presented in Figure 3 for a large set of momentum transfers **Q** from ($Q_h$ 0 0), ($Q_h$ 0 -1), and (1 0 $Q_l$) scans in the reciprocal space at $T$ = 2 K. Except at **Q** = (0.6 0 0) and (1.6 0 -1) corresponding to $\mathbf{k}_1$, no field-induced change of the neutron diffracted intensity is observed at the different probed momentum transfers. In particular, no change of intensity occurs in high field at the harmonics $n\mathbf{k}_1$, where $n \neq 1$ is an integer, including the Brillouin zone center $\mathbf{k}_{ZC}$ = (0 0 0), within an error of about 5-10 % of the induced intensity at $\mathbf{k}_1$ (cf. Supplementary Information). The absence of harmonics supports that the high-field magnetic structure is a spin density wave, i.e., a sine-modulated magnetic structure, in opposition with a squared-modulated structure, where harmonics of the main wavevector would lead to full moments, either up or down, on the U sites. Spin-density-wave modulations of the magnetic moment are generally observed at low temperature in itinerant magnets at zero-field, as chromium [35], or, in a magnetic field, as the heavy-fermion superconductor CeCoIn$_5$ [36] and the low-dimensional magnet Sr$_3$Ru$_2$O$_7$ [37]. However, in the present case, we cannot fully exclude the possibility of a 'special' sequence of squared Ising moments, for which high-order harmonics $n\mathbf{k}_1$ would all have intensities smaller than our error bar, i.e., 5-10 % of the magnetic intensity at $\mathbf{k}_1$. The absence of neutron Bragg peaks at $\mathbf{k}_2$ = (2/3 0 0) and $\mathbf{k}_{ZC}$ is in contradiction with the *'up-up-down'* squared modulation proposed from nuclear magnetic resonance (NMR) measured on URu$_2$Si$_2$ in the field-window 35.5-36.5 T [38]. The NMR spectrum, composed of a central narrow peak and two broad satellites, is not compatible with a single-**k** spin-density wave, which would lead either to a few discrete and thin peaks (if $\mathbf{k}_1$ is commensurate) or to a double-horn line shape (if $\mathbf{k}_1$ is incommensurate) [38]. However, an incommensurability of $\mathbf{k}_1$ (i.e., a slight deviation from (0.6 0 0) compatible with our neutron resolution $\Delta\mathbf{k} \simeq (\pm 0.01$ $\pm 0.05$ $\pm 0.003))$, combined with a multi-**k** character of the spin-density-wave, could induce peculiar features, as a central peak, in the NMR spectrum [39], and could possibly reconcile the NMR and neutron studies. Such multi-**k** structure would imply a conservation of the four-fold symmetry, i.e., of the equivalence between $x$ and $y$, in the high-field spin-density-wave phase, in contrast with the two-fold symmetry proposed in [16],[17] for the low-field hidden-order phase.

**Absence of a ferromagnetic Bragg peak.** Surprisingly, the neutron diffracted intensity at the Brillouin zone center $\mathbf{k}_{ZC}$ is field-invariant while the magnetization $M$ is strongly-enhanced above 35 T [22]. Usually, a magnetic Bragg peak develops at $\mathbf{k}_{ZC}$ and has an intensity $\propto M^2$. From $M \simeq 0.5$-$1$ $\mu_B$/U in the field window 35-39 T, and assuming a magnetic form factor corresponding to localized $f$-electrons (estimated from a low-field study of the magnetic form factor [40]), one expects a magnetic intensity of $\simeq$ 2000-8000 counts/s, in addition to the non-magnetic Bragg peak of $\simeq$ 19000 counts/s, at $\mathbf{Q}$ = (1 0 -1) corresponding to $\mathbf{k}_{ZC}$ (cf. Supplementary Information). However, no such change of intensity is observed here at $\mathbf{Q}$ = (1 0 -1), within an error of 200 counts/s. Several scenarios could explain this unexpected feature. (i) The absence of a magnetic Bragg peak at $\mathbf{k}_{ZC}$ could result from a strongly-reduced magnetic form factor $f_M$ related to itinerant electrons. Since the observation of magnetic Bragg peaks at $\mathbf{Q}$ = (0.6 0 0) and (1.6 0 -1) demonstrates that nearly-localized electrons with $f_M \lesssim 1$ drive the high-field spin-density-wave of wavevector $\mathbf{k}_1$, scenario (i) would imply that the high-field magnetism would be localized, or nearly-localized, at $\mathbf{k}_1$ but itinerant at $\mathbf{k}_{ZC}$. (ii) Alternatively, the absence of magnetic Bragg peak at $\mathbf{k}_{ZC}$ could be explained by field-induced magnetic excitations at $\mathbf{k}_{ZC}$, which would contribute to the enhanced magnetization (cf. Kramers-Kronig relationship [41]), but not to the magnetic Bragg peak at $\mathbf{k}_{ZC}$. (iii) Finally, this absence could also result from a high-field decrease of the nuclear Bragg peak at $\mathbf{Q}$ = (1 0 -1) (related via local stress effects to a change of some atomic positions) which would counteract the increase of magnetic intensity at $\mathbf{k}_{ZC}$. The missing magnetic Bragg intensity at $\mathbf{k}_{ZC}$ constitutes an open question and will merit further studies.

**Relationship with Fermi surface nesting.** In an itinerant picture of magnetism, long-range magnetic ordering with a wavevector $\mathbf{k}_m$ can be related to a partial or complete nesting of two parts of the Fermi surface. The dual localized-itinerant behavior is a central issue in heavy-fermion metals, where the magnetic properties often result from itinerant $f$ electrons. In URu$_2$Si$_2$, 5$f$ electrons hybridize with electrons from the conduction bands and modifications of the Fermi surface and carrier mobility accompany the establishment of the hidden-order phase at the temperature $T_0$ [26],[42],[43],[44],[45]. The itinerancy of the $f$ electrons has already been shown to be a key ingredient for the magnetic properties of this system. Angle-resolved-photoemission spectroscopy (ARPES) and band-structure calculations showed that Fermi surfaces centered on the Γ and Z points are nested with the wavevector $\mathbf{k}_0$ [9],[10],[11],[12],[15]. In Refs. [9],[15], a nesting with the wavevector $\mathbf{k}_1$ between Fermi surface 'petals' around the Γ point has been identified. In Refs. [11],[12], a nesting with the wavevector (0.4 0 0) between these petals and a Fermi surface centered on the Γ point has also been found. In Ref. [14], another picture has been proposed, with a nesting with wavevector $\mathbf{k}_0$ between small Fermi pockets centered on the X points (defined in the first Brillouin zone) and a nesting with wavevector $\mathbf{k}_1$ between Fermi surfaces centered on the Γ and Z points. Despite these nesting of the Fermi surface, no long-range magnetic order has been reported yet in URu$_2$Si$_2$ in its hidden-order phase. Instead of long-range magnetic order, inelastic magnetic excitations at the wavevectors $\mathbf{k}_0$ and $\mathbf{k}_1$ have been observed by neutron scattering in the hidden-order phase [5],[6],[7],[8]. Their relationship with nesting effects and long-range magnetic ordering will be discussed below.

Assuming a high-rank multipolar order with wavevector $\mathbf{k}_0$, which cannot be seen directly by neutron diffraction, band structure calculations resulted in a Fermi surface of the hidden-order state which is very similar to that of the pressure-induced antiferromagnetic state [9], in accordance with experimental observations [19]. Under a high magnetic field applied along **c**, successive Fermi surface modifications occur inside the hidden-order phase and at the transition fields $H_1$, $H_2$, and $H_3$ delimiting the high-field phases [24],[25],[26],[27],[28]. Several Fermi surface instabilities in fields smaller than $\mu_0 H_1 = 35$ T, i.e., inside the hidden-order phase [24],[26],[27],[28], are due to Lifshitz transitions and suggest a progressive modification of Fermi surface nestings. Resulting from the Fermi surface reconstruction at $H_1$, the Fermi surface nesting with $\mathbf{k}_1$ is probably reinforced, leading to the onset of the spin-density wave observed here with the wavevector $\mathbf{k}_1$.

**Magnetic fluctuations as precursor of magnetic ordering.** Figure 4 presents the magnetic phase diagram of $URu_2Si_2$ under pressure and magnetic field, over a wide range of temperatures up to 75 K. At zero-field and ambient pressure, $URu_2Si_2$ is in a correlated paramagnetic regime below $T_\chi^{max}$ = 55 K at which the static magnetic susceptibility $\chi$ is maximal [22]. The correlations are accompanied by a strong and broad magnetic-fluctuation spectrum peaked at the wavevector $\mathbf{k}_1$, which progressively sets at temperatures of the order of $T_\chi^{max}$ [8]. When the system enters in its hidden-order phase, the magnetic fluctuations give rise to sharp inelastic spectra at the wavevectors $\mathbf{k}_0$ and $\mathbf{k}_1$, with the gaps $\Delta E_0 \simeq$ 1.7 meV and $\Delta E_1 \simeq$ 4.5 meV, respectively, at low temperature. Below $T_0$, the intensity of the magnetic fluctuations at $\mathbf{k}_0$ suddenly develops while the intensity of the fluctuations at $\mathbf{k}_1$, as indicated by the static susceptibility $\chi'(\mathbf{k}_1,E=0)$, drops by 20 % [8],[46].

The magnetic fluctuations peaked at the hot wavevectors $\mathbf{k}_0$ and $\mathbf{k}_1$ indicate the proximity of quantum phase transitions associated with long-range ordering with the same wavevectors. Such precursory role of the magnetic fluctuations has been emphasized in the heavy-fermion paramagnet $CeRu_2Si_2$, where doping (La, Rh, or Ge) and magnetic field permit to tune competing quantum instabilities (cf. [47] and Refs. therein). With pressure [20],[21] (as well as Rh-doping [30]), $URu_2Si_2$ can be tuned to antiferromagnetic long-range order with wavevector $\mathbf{k}_0$. The disappearance of the hidden order in $URu_2Si_2$ under pressure coincides with a transfer of weight from magnetic fluctuations with wavevector $\mathbf{k}_0$ into antiferromagnetic order with the same wavevector [48],[49]. Contrary to those with $\mathbf{k}_0$, the fluctuations with wavevector $\mathbf{k}_1$ survive under pressure [48] and their progressive variation may be in relationship with the continuous evolution of $T_\chi^{max}$ [50]. With magnetic field, a spin-density wave is stabilized with the wavevector $\mathbf{k}_1$. Both $T_\chi^{max}$ and $T_0$ vanish in the quantum critical area $\mu_0 H \simeq$ 35-39 T [22] where long-range magnetic ordering sets in. In the polarized regime above $\mu_0 H_3$ = 39 T, from the magnetization higher than 1.4 $\mu_B$/U one can expect a loss of intersite magnetic correlations at $\mathbf{k}_0$ and $\mathbf{k}_1$, the magnetic excitations becoming wavevector-independent, i.e., of local nature. This picture is compatible with the observation in fields up to 17 T that the gap in the excitations at $\mathbf{k}_0$ increases with $H$, while that in the excitations at $\mathbf{k}_1$ remains constant, their extrapolations converging at $\simeq 35$ T [51]. In the critical field window [35 T; 39 T], the enhancement of the Sommerfeld coefficient [22] is the signature of critical fluctuations, presumably of magnetic origin (cf. CeRu$_2$Si$_2$, where critical magnetic fluctuations were reported at the La-doping- and magnetic-field-induced instabilities [52],[53]). However, inelastic neutron scattering under steady fields up to 40 T, which is far beyond the present state of the art (26 T achieved recently at the neutron source in Berlin [54]), would be needed to unveil the high-field evolution of the magnetic fluctuations in URu$_2$Si$_2$.

**Discussion**

We succeeded to detect a high-magnetic-field switch from hidden-order to spin-density-wave phases in URu$_2$Si$_2$. This result illustrates the dual localized and itinerant facets of the 5$f$ electrons in this system. The relationship between Fermi surface nesting and the magnetic wavevector $\mathbf{k}_1$ of the spin-density wave, as well as the cascade of Lifshitz transitions induced by the high-field magnetic polarization [24],[25],[26],[27],[28], show the itinerant character of the 5$f$ electrons. The neutron diffracted intensity with the wavevector $\mathbf{k}_1$ is compatible with the magnetic form factor $f_M$ corresponding to U$^{3+}$ or U$^{4+}$ configurations [40] and highlights the localized character of the 5$f$ electrons. Knowing that a change of the Fermi surface in the polarized regime of URu$_2$Si$_2$ was detected indirectly by Hall effect [55] and directly by quantum oscillations experiments [25], a challenge is to determine whether the missing magnetic Bragg peak at the wavevector $\mathbf{k}_{ZC}$ = (0 0 0) in a high magnetic field could be related to a change in the itinerancy of the $f$-electrons.

A relationship between the long-range magnetic order, the Fermi surface and the magnetic fluctuations has been highlighted. In the hidden-order phase, one may suspect 'inelastic' nestings between electronic band surfaces of different energies, to be connected with the observed gapped magnetic excitations at the wavevectors $\mathbf{k}_0$ and $\mathbf{k}_1$ [5],[6],[7],[8]. These excitations are precursors of the long-range-ordered states induced under magnetic field and pressure. Under a magnetic field, the transformation of the paramagnetic state with gapped magnetic excitations into a spin-density wave with wavevector $\mathbf{k}_1$ might be related to the stabilization of Fermi surface nesting with the wavevector $\mathbf{k}_1$. Under pressure, a modification of the Fermi surface nesting with the wavevector $\mathbf{k}_0$ might also lead to long-range magnetic order with the wavevector $\mathbf{k}_0$. Additional effects, as modifications of the crystal-field, could also play a role in the stabilization of magnetically-ordered phases.

Interestingly, the effect of Rh-doping leads to a decrease of $T_0$ and $T_\chi^{max}$ [30],[31] similar to the magnetic field effect, and to antiferromagnetic order with wavevector $\mathbf{k}_0$ (for Rh-contents $0.02 \leq x \leq 0.03$) similar to the pressure effect. The main wavevector $\mathbf{k}_2 = (2/3\ 0\ 0)$ of the high-field squared magnetic structure of $U(Ru_{0.96}Rh_{0.04})_2Si_2$ [33] is rather close to the wavevector $\mathbf{k}_1 = (0.6\ 0\ 0)$ of the spin density wave of $URu_2Si_2$, confirming a relationship between these two magnetic phases. The differences between these wavevectors might result from subtle differences in the band structures of the two compounds, due to the doping-sensibility of their low-carrier density.

Our observation of a spin density wave in magnetic fields between 35 and 39 T will certainly push to develop models incorporating on equal basis the Fermi surface topology and the magnetic interactions. With the objective to describe competing quantum phase transitions between the hidden-order and long-range-ordered phases, such models will be a basis to solve the hidden order problem. From an instrumental point of view, this work paves the way for neutron magnetic diffraction on weak-moment systems under pulsed magnetic field up to 40 T.

**Methods**

The single crystal of mass $m \simeq 1$ g and dimensions 6*4*4 mm$^3$ used in this study has been grown by the Czochralski technique in a tetra-arc furnace. It has been characterized by X-ray Laue. A sample from the same batch was checked by electrical resistivity, showing a residual resistivity ratio of around 50 and a superconducting temperature $T_c$ of about 1.5 K, indicating its high quality. A slit perpendicular to the basal plane was cut into the sample to reduce the surface exposed to the pulsed field and, thus, self-heating of the sample by eddy currents. A new long-duration (rising time of 23 ms and total time of > 100 ms) and high duty-cycle (40-T, 36-T, and 31-T shots can be repeated every 10, 7, and 5 min, respectively) 40-T conical pulsed magnet developed by the LNCMI-Toulouse has been used. A cryostat specially designed by the ILL-Grenoble for this magnet allowed reaching temperatures down to 2 K, and the diffraction experiment was carried out on the triple-axis spectrometer IN22 (Collaborative Research Group-CEA at the ILL, Grenoble) operated in a double-axis mode with an incident neutron wavelength of 1.2 Å. At each momentum transfer, the field-dependence of the neutron intensities has been extracted by summing the data accumulated over a few tens of pulsed field shots, with either constant time- or constant field-integration windows (the rise and fall of the pulsed field were analyzed separately).

## Acknowledgments

We acknowledge R. Ballou, J. Béard, P. Dalmas de Réotier, H. Ikeda, M.-H. Julien, M. Horvatic, E. Ressouche and G Rikken for useful discussions. Part of this work was funded by the ANR grant Magfins N° ANR-10-0431. H.N. acknowledges KAKENHI 23224009. D.A. acknowledges KAKENHI, 15H05745, 15H05882, 15H05884, 15K21732, 25247055. H.N. and D.A. acknowledge support by ICC-IMR.

## Author contributions

All co-authors conceived this project. D.A. has grown, characterized and prepared the single crystal studied here. J.B. and P.F. designed and developed the pulsed-field magnet. X.T. and E.L.-B. designed and developed a special cryostat for this magnet. W.K., F.D., F.B., K.K., H.N. and L.-P.R. performed the neutron scattering experiment. W.K. and F.D. analyzed the data. W.K., F.D., F.B., J.F. and L.-P.R. wrote the paper, with inputs from the other co-authors.

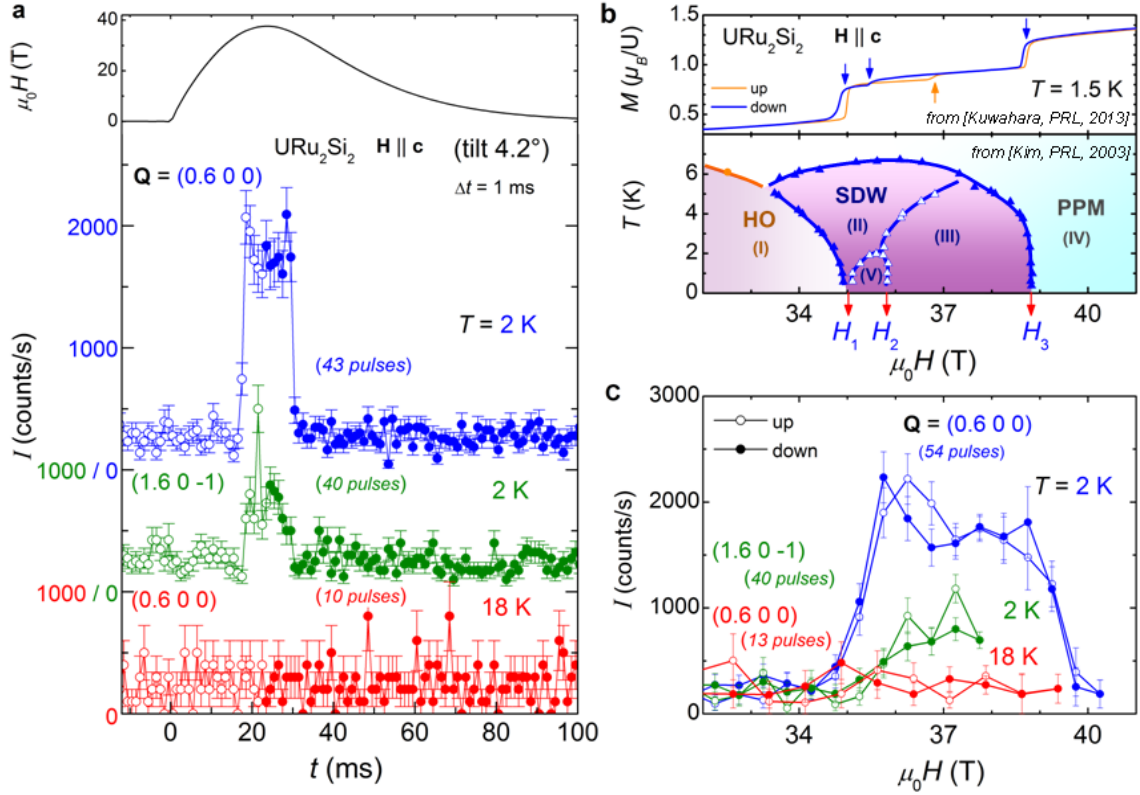

**Figure 1| Diffracted neutron intensity at Q = (0.6 0 0) and (1.6 0 -1), magnetization, and magnetic phase diagram of $URu_2Si_2$ under fields up to 40 T. a.** Time profile of a field pulsed up to 38 T, and time-dependence of the diffracted intensity at **Q** = (0.6 0 0) and $T = 2$ K, **Q** = (1.6 0 -1) and $T = 2$ K, and **Q** = (0.6 0 0) and $T = 18$ K, in fields up to 38 T, integrated within time steps $\Delta t$ = 1 ms. The error bars $\Delta I$ are given by the square root of the neutron counts ($\Delta I = \sqrt{I/\tau}$ where $\tau$ is the total accumulation time over several pulses). **b.** Magnetization versus magnetic field at $T = 1.5$ K (from Ref. [33]) and magnetic-field-temperature phase diagram deduced from magnetoresistivity versus temperature and versus field, in falling fields (from Ref. [23]). **c.** Field-dependence of the same diffracted intensities in fields up to 40.5 T, integrated within field steps $\Delta(\mu_0 H)$ = 0.5 T. Open and full symbols correspond to rising and falling fields, respectively. The magnetic field has been applied along the easy magnetic axis **c** with a tilt of 4.2 ° along **a**.

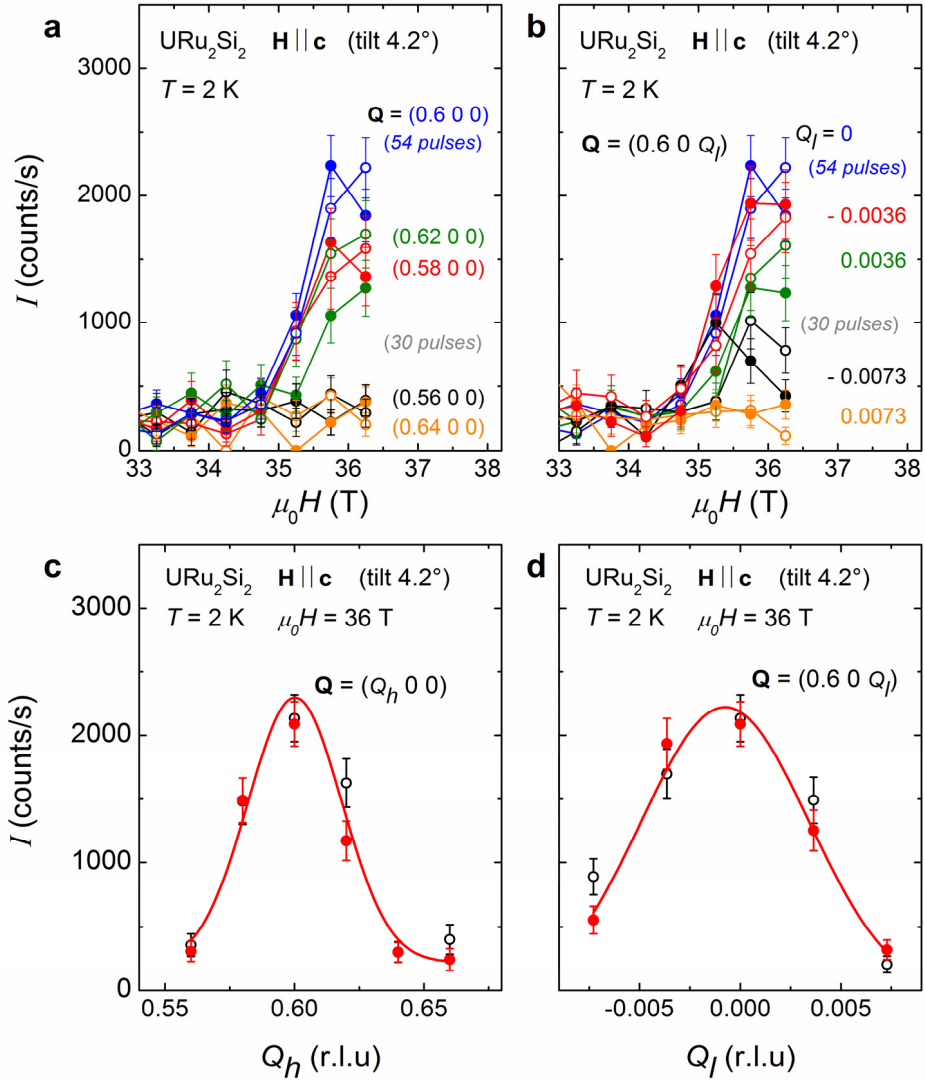

**Figure 2| High-magnetic-field neutron diffracted intensity at momentum transfers close to $Q_1 = (0.6\ 0\ 0)$ at a temperature $T = 2$ K.** Magnetic-field dependence of the neutron diffracted intensity **a.** at momentum transfers $\mathbf{Q} = (Q_h\ 0\ 0)$ and **b.** at momentum transfers $\mathbf{Q} = (0.6\ 0\ Q_l)$, in fields up to 36.2 T, within field steps $\Delta(\mu_0 H)$ of 0.5 T. **c.** $Q_h$ and **d.** $Q_l$ scans around the momentum transfer $\mathbf{Q}_1 = (0.6\ 0\ 0)$ under a magnetic field $\mu_0 H = 36$ T (field window $\Delta(\mu_0 H) = 1$ T). The error bars $\Delta I$ are given by the square root of the neutron counts ($\Delta I = \sqrt{I/\tau}$). Open and full symbols correspond to rising and falling fields, respectively. The magnetic field **H** has been applied along the easy magnetic axis **c** with a tilt of 4.2 ° along **a**.

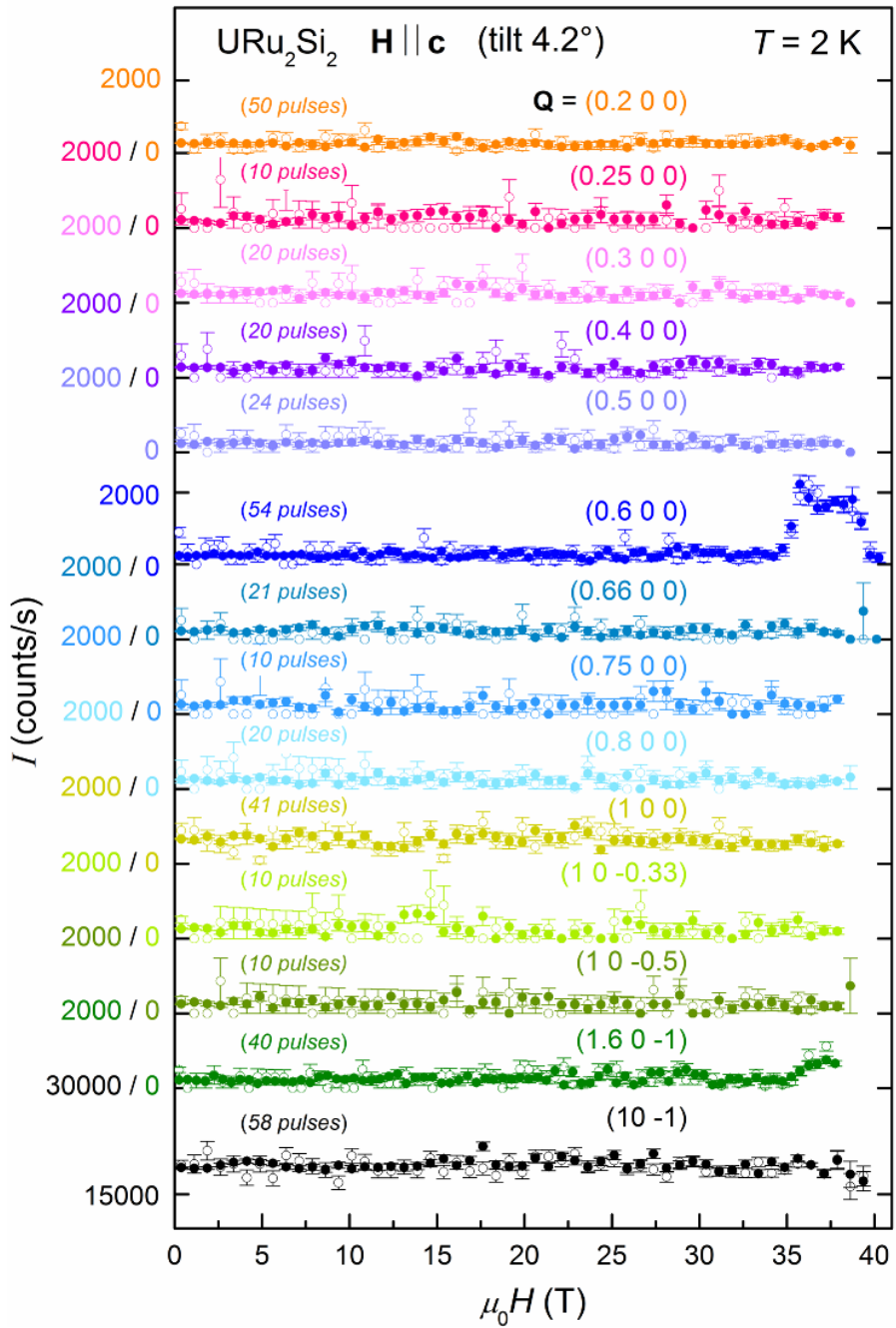

**Figure 3| Magnetic-field dependence of the neutron diffracted intensity for a large set of momentum transfers in fields up to 40.5 T and at $T$ = 2 K.** The error bars $\Delta I$ are given by the square root of the neutron counts ($\Delta I = \sqrt{I/\tau}$). Open and full symbols correspond to rising and falling fields, respectively. The magnetic field **H** has been applied along the easy magnetic axis **c** with a tilt of 4.2 ° along **a**.

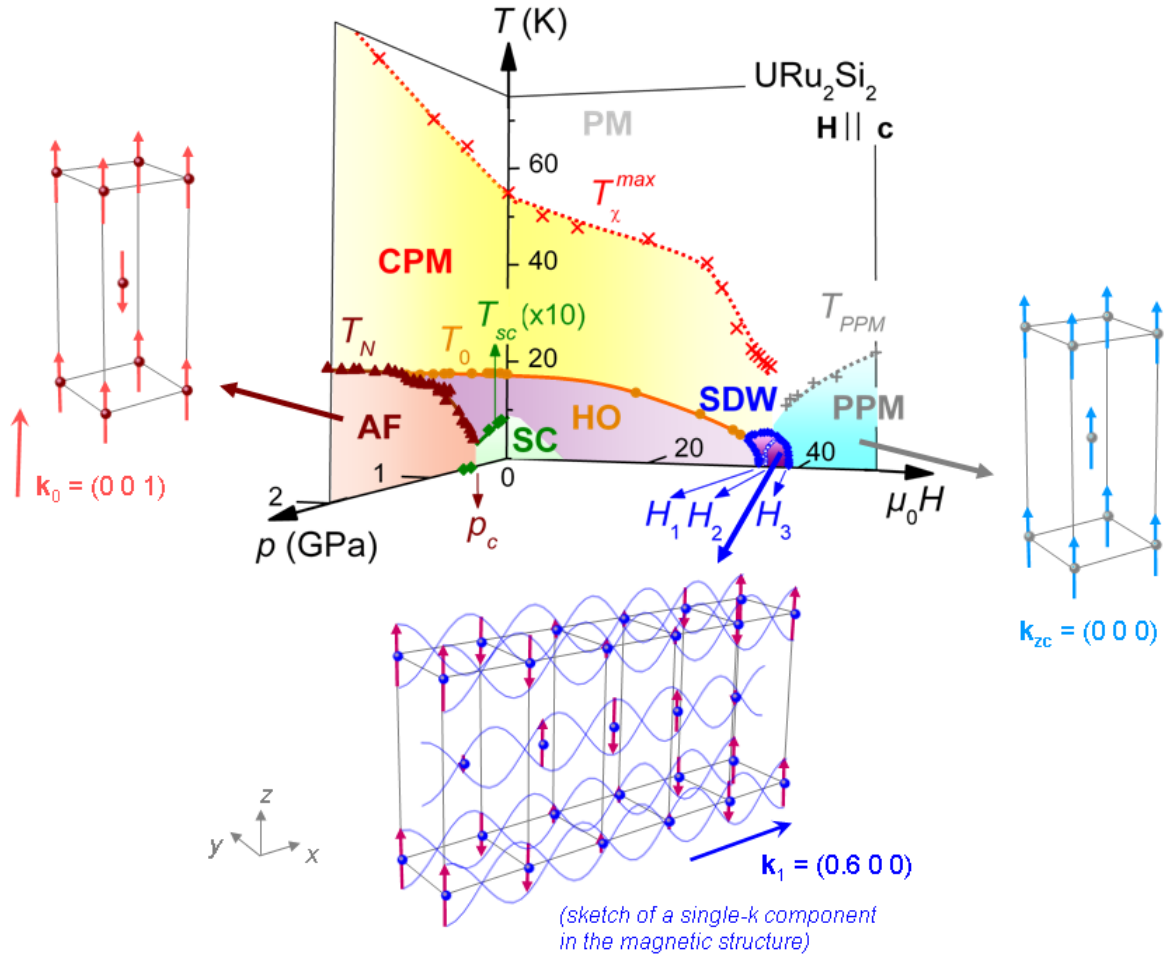

**Figure 4| Temperature-magnetic field and temperature-pressure phase diagrams of $URu_2Si_2$.** PM, CPM, and PPM are the high-temperature, the low-temperature correlated, and the high-field polarized paramagnetic regimes, respectively, HO and SC the hidden-order and superconducting phases, respectively, AF the antiferromagnetic state, and SDW the spin-density wave state. For the AF, SDW, and PPM regimes, a sketch of the magnetic structure is shown, where the spheres correspond to the U ions and the arrows represent their magnetic moments. For clarity, a single-component (wavevector $k_1 = (0.6\ 0\ 0)$) sine-modulation is shown for the SDW, which is suspected to be multi-$k$. Data points in the phase diagram have been compiled from Refs. [19],[22],[23],[31],[50].

**Supplementary Table 1**. Neutron diffracted intensities $I_0$ and $I_m$ summed over the field windows [0;34T] and [35;39T], respectively, at the wavevector $\mathbf{k}_1$ and its main harmonics, and comparison of the field-induced variation of the harmonics intensities with that of $\mathbf{k}_1$. For each momentum transfer $\mathbf{Q}$ are given the magnetic wavevectors $n\mathbf{k}_1$, where $1 \leq n \leq 10$ is an integer, of the contributing harmonics and the corresponding structural Bragg peak $\boldsymbol{\tau}$, verifying $\mathbf{Q} = \boldsymbol{\tau} + n\mathbf{k}_1$ or $\mathbf{Q} = \boldsymbol{\tau} - n\mathbf{k}_1$.

| Momentum transfer $\mathbf{Q}$ | Magnetic wavevector $n\mathbf{k}_1$ | Structural Bragg position $\boldsymbol{\tau}$ | Number of shots | $I_0[0;34T]$ (counts/s) | $I_m[35;39T]$ (counts/s) | $\dfrac{I_m(\mathbf{Q}) - I_0(\mathbf{Q})}{I_m(\mathbf{Q}_1) - I_0(\mathbf{Q}_1)}$ |
|---|---|---|---|---|---|---|
| (0.2 0 0) | $3\mathbf{k}_1$ = (1.8 0 0)<br>$7\mathbf{k}_1$ = (4.2 0 0) | (2 0 0)<br>(-4 0 0) | 50 | 278 ± 6 | 259 ± 18 | (-1.4 ± 1.4) % |
| (0.4 0 0) | $4\mathbf{k}_1$ = (2.4 0 0)<br>$6\mathbf{k}_1$ = (3.6 0 0) | (-2 0 0)<br>(4 0 0) | 20 | 292 ± 10 | 268 ± 29 | (-1.8 ± 2.2) % |
| $\mathbf{Q}_1$ = (0.6 0 0) | $\mathbf{k}_1$ = (0.6 0 0)<br>$9\mathbf{k}_1$ = (5.4 0 0) | (0 0 0)<br>(6 0 0) | 54 | 256 ± 6 | 1627 ± 42 | 1 |
| (0.8 0 0) | $2\mathbf{k}_1$ = (1.2 0 0)<br>$8\mathbf{k}_1$ = (4.8 0 0) | (2 0 0)<br>(-4 0 0) | 20 | 272 ± 9 | 212 ± 25 | (-4.4 ± 2.0) % |
| (1 0 0) | $5\mathbf{k}_1$ = (3 0 0) | (-2 0 0) | 41 | 705 ± 11 | 604 ± 30 | (-7.4 ± 2.4) % |
| (1 0 -1) | $0\mathbf{k}_1$ = (0 0 0)<br>$10\mathbf{k}_1$ = (6 0 0) | (1 0 -1)<br>(-5 0 -1) | 58 | 19138 ± 46 | 19236 ± 156 | (7.2 ± 11.9) % |

## Supplementary Note 1

**Amplitude of the ordered magnetic moment**

Assuming magnetic moments aligned along the easy magnetic axis **c**, the amplitude of the magnetic moment associated with a wavevector $\mathbf{k}_M$ is given by:

$$M(\mathbf{k}_M) = \frac{1}{f_M p |(\sin\alpha)|} \left| \sum_j b_j e^{i\mathbf{Q}_N \cdot \mathbf{r}_j} \right| \sqrt{\frac{I_M / L_M}{I_N / L_N}}, \quad (S1)$$

where $\mathbf{Q}_N$ is the momentum transfer at a nuclear Bragg position, $I_M$ and $I_N$ are the measured intensities (integrated over the sample precession angle $\omega$) at $\mathbf{Q}_M$ (corresponding to the magnetic wavevector $\mathbf{k}_M$) and $\mathbf{Q}_N$, respectively, $L_M$ and $L_N$ are the Lorentz factors at $\mathbf{Q}_M$ and $\mathbf{Q}_N$, respectively, $f_M$ is the magnetic form factor at $\mathbf{Q}_M$, $p = 0.2696.10^{-12}$ cm, $\alpha$ is the angle between the moment and **Q**, and $b_j$ and $\mathbf{r}_j$ are the neutron scattering length and position of the ions $j$, with $b_U = 0.842.10^{-12}$ cm, $b_{Ru} = 0.721.10^{-12}$ cm, and $b_{Si} = 0.4149.10^{-12}$ cm [1].

Here, we use equation (S1), assuming a magnetic form factor corresponding to localized or nearly-localized $f$-electrons (estimated from a low-field study of the magnetic form factor [2]). From the intensity measured (and integrated over $\omega$ scans) at the magnetic Bragg peak at **Q** = (0.6 0 0) and at the nuclear Bragg peak at $\mathbf{Q}_N$ = (1 0 -1), we extract the moment amplitude $M(\mathbf{k}_1)$ = 0.24 ± 0.02 $\mu_B$/U. From the field-induced magnetic Bragg peak at **Q** = (1.6 0 -1), which is equivalent to that measured at **Q** = (0.6 0 0) but whose intensity is roughly four-times smaller (cf. Figure 1), we extract a similar value $M(\mathbf{k}_1)$ = 0.27 ± 0.03 $\mu_B$/U. We also estimate that, for 35 < $\mu_0 H$ < 39 T, an increase of $\simeq$ 2000 - 8000 counts/s would be expected at **Q** = (1 0 -1) if the magnetization of 0.5 - 1 $\mu_B$/U was related to a magnetic Bragg peak at $\mathbf{k}_{ZC}$ = (0 0 0).

## Supplementary Note 2

**Spin density wave vs. squared modulation**

Assuming a multi-**k** structure (made either of harmonics of the main wavevector or of equivalent wavevectors), the moment on the U site of position **r** is given by the superposition of the sine-modulated moments associated with each component $\mathbf{k}_i$:

$$M_U(\mathbf{r}) = \sum_{i=1}^{N} M_U(\mathbf{k}_i, \mathbf{r}), \quad (S2)$$

where N is the number of considered wavevectors,

$M_U(\mathbf{k}_i, \mathbf{r}) = M(\mathbf{k}_i)\cos(2\pi\mathbf{k}_i\mathbf{r} + \varphi_i)$ if $\mathbf{k}_i$ is at the Brillouin zone center,

$M_U(\mathbf{k}_i, \mathbf{r}) = 2M(\mathbf{k}_i)\cos(2\pi\mathbf{k}_i\mathbf{r} + \varphi_i)/N$ if $\mathbf{k}_i$ is at a Brillouin zone border shared by $N$ zones, and $M_U(\mathbf{k}_i, \mathbf{r}) = 2M(\mathbf{k}_i)\cos(2\pi\mathbf{k}_i\mathbf{r} + \varphi_i)$ for all other $\mathbf{k}_i$.

Supplementary Table 1 presents, for several momentum transfers **Q** corresponding to the harmonics $n\mathbf{k_1}$, the neutron intensities $I_0$ integrated in the field window [0;34T] in the hidden-order phase and $I_m$ integrated in the field window [35;39T] in the field-induced ordered phase, resulting from summations over 20 to 58 pulsed fields shots. The ratios between [$I_m(\mathbf{Q}) - I_0(\mathbf{Q})$] and [$I_m(\mathbf{Q}_1) - I_0(\mathbf{Q}_1)$], which are the field-induced variations of intensity at **Q** and $\mathbf{Q}_1$, respectively, indicate that no magnetic Bragg peak develops at the harmonics in high magnetic field, within an error of 5-10 % of the magnetic intensity at $\mathbf{k}_1$ (5 % error or less for **Q** = (0.2 0 0), (0.4 0 0), (0.8 0 0), and (1 0 0), and 12 % error for **Q** = (1 0 -1)). The non-observation of harmonics within these experimental errors permits to rule out the picture of a squared magnetic structure of URu$_2$Si$_2$ between 35 and 39 T and, thus, to support the picture of a spin-density wave with the wavevector $\mathbf{k}_1$ = (0.6 0 0).

## Supplementary references